# News from the Internet congestion control world

Dario Rossi, Claudio Testa, Silvio Valenti, Paolo Veglia
TELECOM ParisTech
Paris, France
firstname.lastname@enst.fr

Luca Muscariello
Orange Labs
Paris, France
luca.muscariello@orange-ftgroup.com

*Abstract*—A few months ago, the BitTorrent developers announced that the transfer of torrent data in the official client was about to switch to uTP, an application-layer congestion-control protocol using UDP at the transport-layer. This announcement immediately raised an unmotivated buzz about a new, imminent congestion collapse of the whole Internet. Though this reaction was not built on solid technical foundation, nevertheless a legitimate question remains: i.e., whether this novel algorithm is a necessary building block for future Internet applications, or whether it may result in an umpteenth addition to the already well populated world of Internet congestion control algorithms.

In this paper, we tackle precisely this issue. The novel protocol is now under discussion at the IETF LEDBAT working group, and has been defined in a draft document in March 2009, whose adoption decision will be taken at the beginning of August 2009. Adhering to the IETF draft definition, we implement the LEDBAT congestion control algorithm and investigate its performance by means of packet-level simulations. Considering a simple bottleneck scenario where LEDBAT competes against either TCP or other LEDBAT flows, we evaluate the fairness of the resource share as well as its efficiency. Our preliminary results show that indeed, there is an undoubted appeal behind the novel application-layer congestion-control protocol. Yet, care must be taken in order to ensure that some important points, such as intra-protocol fairness, are fully clarified in the draft specification – which we hope that this work can contribute to.

## I. Introduction

A few months ago, a post in the thread announcing the new $\mu$Torrent release 1.9 alpha 13485 in the BitTorrent developer forum [1] raised a lot of motivated interest as well as quite a few unmotivated buzz [2]–[5]. The main novelties consisted in the fact that i) starting from the new release, the official BitTorrent client would no longer be open-source but closed and proprietary, and that ii) data download would now use a new protocol. The new protocol, named "micro transport protocol" (uTP), was described as an application-layer protocol for data transfer, implementing a novel congestion-control algorithm built on top of UDP at the transport layer.

Nevertheless, the main item retained was that BitTorrent would have switched its data transfer over UDP – which do not implement any kind of congestion control and is thus usually associated with *unresponsive* source. This fallacious interpretation raised serious concerns: as BitTorrent constitutes a significant portion of nowadays Internet traffic, its switchover to UDP was seen as the cause for the forthcoming collapse of the network [2]. This "Internet meltdown" buzz rapidly flooded to popular websites [3], and only after an official reaction of BitTorrent followed by intense discussions, this climax started slowing down [4], [5].

Yet, the buzz was not built on solid technical foundation. In fact the original post [1] clearly states: *"This UDP-based reliable transport is designed to minimize latency, but still maximize bandwidth when the latency is not excessive. In addition, [...] uTorrent, when using uTP, should not kill your net connection – even if you do not set any rate limits."* Developers goal was to build a protocol able to *"detect problems very quickly and throttle back accordingly so that BitTorrent doesn't slow down the Internet connection and Gamers and VoIP users don't notice any problems."* Finally they affirm that *"uTP is the result of a couple of years of work to try to make a BitTorrent protocol that works better on the Internet [...] trying to do our bit to be responsible citizens on the Internet"*, also pointing out the co-chairing effort of an IETF working group on Low Extra Delay Background Transport (LEDBAT) [6], whose first draft [7] dates March 2009. Thus, as openly discussed at IETF [8], the BitTorrent position clearly goes in the direction of ISP-friendliness (for what concerns an AS-aware peer selection process) and TCP-friendliness (for what concerns the congestion control mechanism employed for the data transfer).

The novel congestion control algorithm is described in [7]. LEDBAT is a windowed protocol, governed by a linear controller designed to infer earlier than TCP the occurrence of congestion on a network path. LEDBAT congestion control is based on the estimation of one-way delay: queuing delay is estimated as the difference between the instantaneous delay and a base delay, taken as the minimum delay over the previous observations. Whenever a sender estimates that the one-way delay is growing, it infers that queue is building up and reacts by decreasing its sending rate. This way, it reacts earlier than TCP, which instead has to wait for a packet loss event to infer that congestion occurred.

While the LEDBAT design goals are sound, and results in [8] state that simulation and large-scale experiments have yielded good results, technical points have been raised by the scientific community participating to the LEDBAT working group, that ongoing discussion has not fully flattened yet [9]. A legitimate question is whether the novel LEDBAT [7] addition to the already well populated world of Internet congestion control algorithms is really necessary and motivated, or whether it would be better instead to rely on already existing, and therefore more stable and better understood, algorithms. These facts, coupled with the move toward a closed and proprietary code, motivates the need for independent studies, so that claims concerning, e.g., the friendliness and efficiency of this new protocol, can be confirmed by independent research.

This work tackles precisely this issue, using event driven packet-level simulations to assess the performance of the LEDBAT controller. Our aim is not to propose any modification to LEDBAT: rather, we aim at evaluating the draft specification [7] as is. Therefore, strictly adhering to the specification, we implement and evaluate the simplest controller that satisfy all the drafts requirement in ns2 [10]. The source code of our LEDBAT implementation is made available to the scientific community upon request.

To summarize our main results, we find that the *linear controller is enough to achieve inter-protocol fairness*: in other words, LEDBAT does not interfere with CBR VoIP/Gaming flows, and guarantees to TCP a more-than-friendly share of bottleneck resource. Moreover, TCP friendliness on a fair basis (i.e., equal competition for resources) is guaranteed also in case of wrong parameter settings. Concerning the link utilization, LEDBAT resource usage is more efficient than TCP whenever the former is alone on the bottleneck. In case both LEDBAT and TCP are present on the link, the overall link utilization increases, since LEDBAT is able to use the resources available beyond those used by TCP, while at the same time not interfering with TCP AIMD dynamics (i.e., LEDBAT reacts earlier than TCP).

Yet, we also find that the *linear controller alone may not solve the issue concerning intra-protocol fairness*, i.e., fairness among competing LEDBAT flows. More precisely, as feared in [9], a late-comer advantage may arise, where newly born connections may absorb all resources, bringing already started connections to starvation. Basically, unfairness is due to an incorrect estimation of the base delay as performed by late-comer connections, which in turns yields new-comers to underestimate the actual queuing delay.

Interestingly, we show that, a *slow-start phase in needed in order to break unfair situation in which the linear controller may get stuck*. Intuitively, slow-start induces losses on already active connections, allowing the capacity to drain the queue empty, so that all connection can get a correct estimate of the base delay. Also, as losses has to be induced only at the beginning of the connection, their impact on CBR VoIP/Gaming flows is likely to be negligible, although a more careful analysis is needed in this context. Therefore, it seems that slow-start is necessary for its side effect on fairness issues, rather than for efficiency matters. At the same time, we point out that the slow-start phase is only specified to be optional in [7]: thus, the above observations suggest that slow-start should be a *mandatory* component of the novel protocol.

Overall, LEDBAT has an undoubted appeal to become a very useful Internet building block: yet, we underline that whether this will happen, however not only depends on its network friendliness (i.e., which helps relieving congestion on user access links), but also on the overall performance of applications relying on it (e.g., BitTorrent download time), as this will have a major impact on users and their consensus.

## II. LEDBAT OVERVIEW

This section provides a basic overview of the LEDBAT draft [7]. To better understand the motivations behind LEDBAT, let us recall that the standard TCP congestion control mechanism needs losses to back off: thus, under a drop-tail FIFO queuing discipline, this means that TCP necessarily fills the buffer. As uplink devices of low-capacity home access networks can buffer up to hundreds of milliseconds [9], this may translate into poor performance of interactive applications (e.g., slow Web browsing and bad gaming/VoIP quality).

To avoid this substantial drawback, LEDBAT implements a distributed congestion control mechanism, tailored for the transport of non-interactive traffic with lower than Best Effort (i.e., TCP) priority, whose main design goals are:

- Saturate the bottleneck when no other traffic is present, but quickly yield to TCP and other UDP real-time traffic sharing the same bottleneck queue.
- Keep delay low when no other traffic is present, and add little to the queuing delays induced by TCP traffic.
- Operate well in drop-tail FIFO networks, but use explicit congestion notification (e.g., ECN) where available.

Intuitively, to saturate the bottleneck it is necessary that queue builds up: otherwise, when the queue is empty, at least sometimes no data is being transmitted and the link is under-exploited. At the same time, in order to operate friendly toward interactive applications, the queuing delay needs to be as low as possible: LEDBAT is therefore designed to introduce a non-zero *target* queuing delay.

To this extent, the LEDBAT controller exploits the ongoing data transfer to perform *one-way delay measurement*, by timestamping packets. One-way delay is used instead of round-trip delay, so that unrelated traffic on the reverse path does not interfere with the data transmission. LEDBAT controllers then estimate the *queuing delay* as the difference between the current delay and a *base delay*, taken as the minimum delay over a number of previous observations.

Finally, LEDBAT adapts its sending rate with a *linear controller*, aiming to keep the estimated queuing delay equal to its target. Also, in order to be TCP-friendly, the controller is designed in such a way that the ramp-up of the congestion window is not higher than that of TCP during congestion avoidance, and that reaction to losses is the same as TCP. In the reminder of this section, we introduce the LEDBAT pseudocode, and report considerations concerning the linear controller, the delay measurement and the TCP-friendliness issues.

### A. LEDBAT Operations

For the sake of simplicity, we consider a bidirectional LEDBAT communication between a sender, having unlimited amount of data to send, and a receiver, merely acknowledging each received data packet. We consider data packets of fixed size. As the draft specifies, LEDBAT can implement a TCP-like slow-start behavior, but *"conservative implementations*

```
on data_packet @ RX:
    remote_timestamp = data_packet.timestamp
    acknowledgement.delay =
        local_timestamp() - remote_timestamp

on acknowledgement @ TX:
    current_delay = acknowledgement.delay
    base_delay = min(base_delay, current_delay)
    queuing_delay = current_delay - base_delay
    off_target = TARGET - queuing_delay
    cwnd += GAIN * off_target / cwnd
```

Fig. 1. Pseudocode of the LEDBAT sender and receiver operations

*MAY skip slow-start altogether"* [7]. For the time being, we thus neglect the slow-start phase.

To perform one-way delay measurements, each data packet contains a header field timestamp: the sender puts a timestamp from its clock into this field. Also, each acknowledgement packet contains a delay field, that the receiver sets to the difference between its local timestamp and the remote timestamp of the sender. A minimum of the measured delay is maintained by the sender in order to estimate the instantaneous queuing delay, which is then used to modulate the congestion window size. Thus, LEDBAT operations can be simply stated as in Fig. 1, which reports the simplified pseudocode with the same notation of [7]. Notice that the behavior of LEDBAT further depends on two parameters, namely TARGET and GAIN. Quoting the draft specifications *"TARGET parameter MUST be set to 25 milliseconds and GAIN MUST be set so that max ramp up rate is the same as for TCP."*

As far as the setting of the GAIN parameter is concerned, we set it to GAIN=1/TARGET: the reason of our choice will be clarified later on in this section. Concerning instead the queuing delay target parameter, the choice TARGET= 25 ms is primarily motivated by implementation issues: indeed, since the queuing delay has to be inferred by measurement, the delay target should not be smaller than the OSes accuracy in timestamping packet. At the same time, we point out that the mandatory choice of a *constant* and furthermore *specific value* for its setting has been largely debated on the WG [9], but consensus has not been reached yet (e.g., unfairness issue may arise in case of non-compliant implementations using different targets). For the time being, we adhere to the draft specification and leave this issue for further work.

### B. Linear Controller

A *linear controller* governs the dynamic of the congestion window in both the ramp-up and ramp-down phases. The linear controller adapt the window to the estimated delay, thus prior that congestion occurs and packets get lost.

Clearly, when the estimate of the queuing delay is lower than the target (i.e., off_target<0) the sending rate has to increase, so that queuing delay reaches the target. Conversely, when the queuing delay estimate is higher than the target (i.e., off_target>0) the controller slow down the sending rate.

Notice that, in the linear controller, the window growth is proportional to the difference between the queuing delay estimate and the target off_target, multiplied by the GAIN factor. This implies that, growth (or shrink) of the window will be slower as the target is approached and faster when the estimate is far from the target. This is a desirable property: indeed, to avoid oscillations on round-trip-time scale, the response of the controller needs to be near zero when the queuing delay estimate is near the target; similarly, to converge faster, the controller response needs to increase faster as the offset from the target increases.

### C. One-way Delay Estimate

As previously stated, one-way delay measurements are performed by adding a timestamp to the packets on the data direction and a measurement result field on the acknowledgement direction. One-way delay is notoriously difficult to measure in the Internet by non-synchronized hosts: however, LEDBAT does not need an accurate estimate of the one-way delay, but only of its *variation* with respect to a *base delay*.

Consider that one-way delay is caused by different components: namely, propagation, transmission, processing and queuing. Neglecting the processing delay, propagation and transmission delays are the only constant components, while the only variable component is constituted by the queuing delay. As can be seen from Fig. 1, the base_delay is continuously updated so as to store the *minimum* delay over all observations: intuitively, packets that will find the queue empty (i.e., null queuing delay) will yield an accurate estimate of the constant portion of the one-way delay (i.e., the sum of propagation and transmission delays). Then, as queuing delay is always non-negative, it can be estimated as the difference between the current and the base delay.

We now argue if and how the queuing delay estimate is affected by timestamp errors (such as fixed offsets from the true time and skews). Concerning the sender and receiver offsets, it is easy to gather that, though the clock offset affects the absolute one-way delay estimate, it however cancels in the arithmetic difference operation queuing_delay = current_delay - base_delay (since both delays are computed as the difference of the receiver minus the sender delay in their turn). Therefore, no synchronization is necessary between peers wishing to use LEDBAT for data transport. Similar considerations, that we are unable to report here for lack of space, about clock skew, noise filtering and route changes issues are addressed in [7], to which we refer the reader for further details. We point out that our ns2 implementation supports all mandatory LEDBAT features, including those needed to cope with route changes on long timescales. However, for the purpose of clarity, in this paper we focus on shorter timescales to avoid this level of detail.

### D. TCP Friendliness Consideration

An important goal of LEDBAT concerns its ability to yield to TCP traffic when sharing the same bottleneck resources. This means that LEDBAT should be able to detect the traffic already present on links, as well as to yield quickly to newly incoming connections (releasing resources such as occupied buffer space and capacity). At the same time, LEDBAT must avoid starvation: indeed, while it is desirable for LEDBAT to quickly yield in presence of interactive traffic such as short Web or Mail transfers, it could be reasonable to compete in a more aggressive fashion with a long-lived FTP transfer. This is an important point: though the right fairness balance might be subjective (depending on the relative importance users attach to, e.g., P2P, Web, VoIP, etc.) in case LEDBAT would always unconditionally yield to any traffic, users could possibly simply revert to TCP based transfers.

A first necessary condition for TCP friendliness, is that LEDBAT *should never ramp-up faster than TCP*. Since the maximum speed with which LEDBAT can increase its congestion window is when the queuing delay estimate is zero (in reason of our earlier observation on the linear controller), it is sufficient to limit this ramp-up speed to match that of TCP in congestion avoidance (i.e., one packet per RTT). Moreover, since delay estimate is always non-negative, this will ensure never ramping-up faster than TCP would (as the TCP ramp-up speed is only attained when no queuing occurs). Notice that our choice of GAIN=1/TARGET satisfies this constraint, since the window growth equals one packet per RTT when the queuing delay is null.

A second necessary condition is that one-way delay based LEDBAT congestion controller *should react early that loss-based TCP controller*: intuitively, if LEDBAT can ramp-down faster than loss-based connection ramps-up, LEDBAT will yield. As early observed, LEDBAT ramps-down when queuing delay estimate exceeds the target and, the more the excess, the faster the ramp-down. The draft states that LEDBAT should *"yield at precisely the same rate as TCP is ramping-up when the queuing delay is double the target"*. Notice that our choice of GAIN=1/TARGET also satisfies this constraint: when the queuing delay is twice the target, it is easy to gather that LEDBAT will ramp-down at a rate equal to one packet per RTT, matching thus TCP congestion avoidance ramp-up speed.

A third necessary condition is that, *in case of loss, LEDBAT should behave like TCP does*. This means that, in case a loss event is detected, LEDBAT will halve its congestion window (halving may happen at most once per RTT). Notice also that, in case of wrong queuing delay estimates that correspond to the most aggressive LEDBAT behavior (i.e., when queuing delay is always estimated to be null), LEDBAT degenerates into a TCP-like behavior, as it will ramp-up as fast as TCP and halve its rate in case of loss. This not only ensure protection against severe congestion (i.e., when most packets are lost) but also results in a conservative approach in case of incorrect queuing delay estimation.

### III. SIMULATION PRELIMINARIES

#### A. Reference scenario

As reference scenario, we consider a bottleneck link of capacity $C$ Mbps and buffer size $B$ packets. For the sake of simplicity, we consider that all transceivers adopt $P = 1500$ Bytes fixed-size packets. Traffic flows in a single direction, and acks are not delayed, dropped nor affected by cross-traffic on their return path. In the following, we consider only homogeneous settings in which all flows have the same round trip time $RTT = 50\,ms$, half of which is due to the propagation and transmission delay components of the bottleneck link (i.e., a one-way base delay of 25 ms).

We devise two different access scenarios, namely ADSL and high-speed (HS). We set ADSL downlink/uplink capacity to $C = 2$ Mbps and 500 Kbps, while we consider a symmetric link of $C = 10$ Mbps capacity in both directions for the HS scenario. Given our round trip time choice, we notice that the bandwidth delay product is equals to 12500 Bytes (8.3 packets) in the ADSL case and 62500 Bytes (41.6 packets) in the high-capacity case. We consider different buffer sizes in $B \in [10, 100] \subset \mathbb{N}$ packet, and notice that a buffer size slightly above the bandwidth delay product is met when $B^\star_{ADSL} = 10$ and $B^\star_{HS} = 50$ packets.

Notice that, having fixed the link capacity $C$ (and the packet size $P$), we can express the queuing delay TARGET in terms of either a time-lapse or bytes (and packets). Denoting for short the TARGET as $\tau$, in the following we will refer indifferently to the queuing delay expressed in terms of time-lapse $\tau_T = 25$ ms or packets $\tau_P = \tau_T C/8P$ (where we assume capacity to be expressed in Mbps and packet size in Bytes). Notice that, in the ADSL scenario, $\tau_T = 25$ ms corresponds to a $\tau_{P,ADSL} = 4.2$ packets, while it corresponds to $\tau_{P,HS} = 20.8$ packets in the HS case. Thus, in both scenarios, buffer sizes $B^\star_{ADSL}$ and $B^\star_{HS}$ can accommodate twice as much queuing delay than the LEDBAT target $\tau$.

#### B. Implementation details

To avoid dealing with the complexity of retransmission in case of loss, we implement our LEDBAT controller as a novel flavor of TCP, of which we change the congestion control mechanism. More precisely, we turn off all TCP feature (e.g., FastRetransmit), leaving only the congestion control algorithm early described in Sec. II. For timestamping purposes, we exploit the TCP timestamping option [12].

We implement all mandatory as well as optional features of LEDBAT [7]. More precisely, we implement a cache of queuing delay minima, mandatory to cope with route changes on long timescales. As far as the optional slow-start phase is concerned, since the LEDBAT draft lacks its description [7], we adopt the standard TCP mechanism. However, unless otherwise stated, slow-start mechanism is turned off. Also, though this issue is not treated in [7], our LEDBAT implementation can work in batch-mode (i.e., all packets of a window are possibly sent out in bursts) or paced-mode (i.e., delaying the packet transmission so that packets are spaced equally during the RTT). Unless otherwise stated, packet pacing is turned on.

Then, notice that reducing the sending window to 0 constitutes a problem, since the linear controller will no longer be able to get one way delay estimates – thus, it will not be able to ever increase its sending window again. Therefore, we set a congestion window minimum of 1 packet per RTT, although this is not explicitly specified in [7].

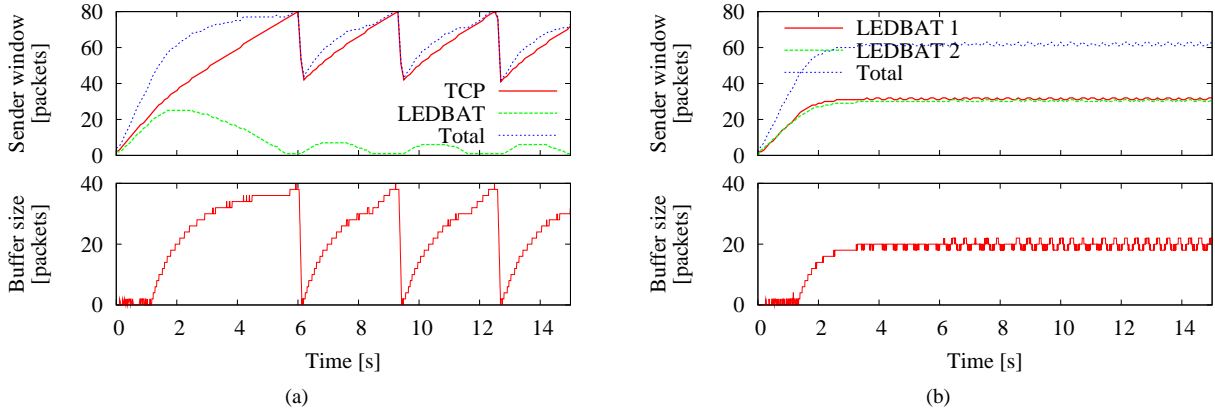

Fig. 2. Temporal evolution of the sender window (top) and of the queue size (bottom) for TCP-LEDBAT (a) and LEDBAT-LEDBAT interaction (b)

Finally, we point out that we built LEDBAT using the `tcp-linux` module, which allows to bridge real Linux code directly into the simulator. As a non-negligible side-advantage, the implementation is then available as a kernel module offering a novel transport-layer protocol that can be used by (unmodified) real applications. However, for the time being we limit our evaluation of LEDBAT performance to simulation, leaving testbed experimentation for future work.

## IV. SIMULATION RESULTS

In this section, we report results gathered with our implementation of the LEDBAT controller in the Network Simulator `ns2` [10]: we start by illustrating some telling examples of the LEDBAT dynamics in simple cases, incrementally adding complexity to refine the picture later on.

As performance metrics, we consider the *fairness* and *efficiency* of the data transfer. For the former, we use Jain's fairness index $F$, which is defined as [11]:

$$F = \frac{(\sum_{i=1}^{N} x_i)^2}{N \cdot \sum_{i=1}^{N} x_i^2} \qquad (1)$$

where $\{x_i\}_{i=1}^{N}$ is the set of rates achieved by $N$ flows sharing the same bottleneck resource. In the best case where each of the $N$ flows gets the same bottleneck share (i.e., $x_i = C/N, \forall i$), fairness is equal to 1, while it decreases to $1/N$ in the worst case where a single flow $j$ takes over the resource forcing others to starvation (i.e., $x_i = 0, \forall i \neq j$). Notice that, since LEDBAT aims at offering a *lower* than best-effort traffic, we expect fairness measure to be lower than 1 when the bottleneck is shared with TCP traffic (inter-protocol fairness) but we expect it to be close to 1 when only LEDBAT flows share the same bottleneck (intra-protocol fairness). As a measure of efficiency, we consider the *link utilization* $\eta$, defined as the ratio of the overall link throughput normalized over the link capacity $C$ (link utilization is evaluated at IP layer, and thus includes the L3 and L4 headers along with the payload).

### A. Ideal Case: Homogeneous Initial Conditions

Let us start our investigation by considering the case where a LEDBAT flow competes for the same bottleneck resources with either i) a TCP flow or ii) another LEDBAT flow. For the time being, let us consider the ideal case where neither LEDBAT nor TCP implement the slow-start phase: in other words, we are interested in observing the coexistence of a LEDBAT linear controller with a TCP AIMD controller, and in evaluating their mutual influence on the congestion window dynamics. Moreover, we assume that both flows start at time $t = 0$ (i.e., homogeneous conditions), when the queue is empty and no other traffic is present on the link. Given these initial conditions, LEDBAT flows are able to measure at $t = 0$ a base delay which, as the queue is empty, is a good estimate of the propagation plus transmission delay components.

Fig. 2-(a) shows the temporal evolution of the LEDBAT and TCP window (top) as well as the queue size (bottom), when $C = 10$ Mbps and $B = 40$ packets (notice that similar qualitative behavior can be obtained on the ADSL scenario as well). From the picture, one can recognize the usual TCP sawtooth behavior, and identify a number of cycles. During the initial ramp-up ($t < 2$ s), LEDBAT and TCP windows grow *nearly* at the same speed of one packet per RTT. Indeed, consider that LEDBAT growth is maximum when the queue length estimates is zero: this happens at the beginning of the simulation, where the link has spare capacity to serve incoming packets, and where both LEDBAT and TCP increments their window by one packet unit per RTT. Then, due to its continuous evaluation of the one way delay, as soon as queue starts to build up LEDBAT senses a growing delay: the linear controller reacts accordingly, slowing down the ramp-up with respect to TCP (which still increments its window by one packet unit per RTT).

Soon after $t = 2$ s, LEDBAT hits the target of $\tau_{P,HS} = 20.8$ packet, and stops the window growth (as it can be seen by the flatness of the sender window curve). When $t > 2$ s, the queue continues to grow until the estimated queuing delay exceeds the target: the controller thus responds to the growing queuing delay by decreasing its window (unlike TCP). The decrease of the LEDBAT window continues until it reaches its minimum sending rate, slightly before $t = 6$ s. TCP instead continues its additive increase until, slightly after $t = 6$ s, it causes the buffer to overflow, halving its window (to about 40 packets) as a consequence.

Soon after the loss event, a new cycle begins, with TCP beginning to increase its window again. However, since TCP abruptly drops its window, the capacity drains the queue empty: given the minimum sending rate of one packet per RTT, LEDBAT has the chance to measure a queuing delay reduction, to which it reacts by opening its window. However, the TCP window at the beginning of the new cycle is no longer starting from 0, but from about 40 packets: therefore, TCP is able to create queuing sooner with respect to the first cycle. As a result, in the second cycle, LEDBAT window growth is slower than during the first one. Moreover, as TCP immediately contributes to queuing, the LEDBAT offset from the target diminishes, and so does its window growth rate. Consequently, LEDBAT growth also stops earlier than before (at about $t = 7$ s), with TCP occupying now a larger buffer portion with respect to the previous cycle. As LEDBAT window temporary settles to a lower window value, the window shrink phase is also shorter (ending soon after $t = 7$ s). TCP is then alone in the link and pushes its window to grow until a loss happens (then, another cycle begins: notice that subsequent cycles are similar to the second).

The dynamics shown in Fig. 2-(a) work as expected: LEDBAT is able to react *earlier* than TCP by estimating the queuing delay, and yielding to TCP, which is able to *work undisturbed*: notice indeed that losses are due to the normal AIMD dynamic of TCP rather than by the LEDBAT-TCP interaction. In the case of figure, the fairness equals $F = 0.65$, with TCP transferring 6 times as much data with respect to LEDBAT during the same time-frame. Fig. 2-(a) also show the sum of both TCP and LEDBAT sender windows, which can be thought an estimate of the instantaneous link utilization: interestingly, notice that during the time period where TCP and LEDBAT coexist on the link, its *utilization increases* with respect to the case where TCP is alone. In the case of figure, the utilization increases by 16%, compared to the case where TCP is alone on the bottleneck, and by 28% compared to the case where two ideal TCP AIMD sources share the bottleneck.

The similar case in which two LEDBAT sources start competing, at $t = 0$ for the bottleneck resources, is shown instead in Fig. 2-(b), again for $C = 10$ Mbps and $B = 40$ packets. In this case, both sources adopt a linear controller and are able to share resources fairly ($F > 0.99$) and efficiently (efficiency is only 0.7% less than in the Fig. 2-(a) case). As expected once the delay target is reached, the LEDBAT sources settle (since the offset from the target is zero, and so the controller response). Buffer occupancy is also smoother, partly due to the fact that LEDBAT sources adopt pacing. Notice also that, since the two sources started together, they measured the same base delay at $t = 0$. Therefore, whenever each of the source senses that the queuing delay has grown to a value equal to the target, it settles: this happens for each source independently, and each source is thus responsible only for about half of the queuing delay in this case.

### B. Ideal Case: Heterogeneous Initial Conditions

By heterogeneous conditions, we mean different start time (or, equivalently, different initial rates) for different sources. This implies that, in this case, the base delay is not necessarily equal for all sources, meaning that the queuing delay estimate will no longer be the same either. Indeed, assume that the first flow starts at time $t_1 = 0$, while the second flow starts at time $t_2 = t_1 + \Delta T$. In case the queuing delay at $t_2$ is not zero but equal to $t_Q(t_2)$, the second source will over-estimate the base delay $t_B(t_2)$ with respect to the one measured by the first source as $t_B(t_2) = t_B(t_1) + t_Q(t_2)$. So, the second source will set its target to a value higher than the first one, increasing the chances of a buffer overflow.

In case of interaction between LEDBAT and TCP, heterogeneity of initial conditions has a negligible impact. To convince of this, consider that, whenever LEDBAT starts first, it is able to correctly estimate the base delay, so it will yield to TCP. Therefore, problems may arise only whenever the LEDBAT flows starts later at $t_2$, in which case it will over-estimate the base delay (by the amount of TCP packets occupying the buffer at $t_2$). This will in turn make LEDBAT under-estimate the amount of queued packets, thus ending up injecting more packets and *anticipating* the first loss cycle. Thus, by recalling Fig. 2-(a), we have that the TCP sender window at the end of the first cycle will be potentially lower when $\Delta T > 0$ with respect to the homogeneous case $\Delta T = 0$. However, notice that after the loss event the capacity drains the queue, so that LEDBAT will have the chance to correct its faulty estimate of the base delay: thus, queuing delay will not be under-estimated during the second cycle. This implies that LEDBAT will yield to TCP and that TCP window growth will be unaffected from the third cycle onward.

The interaction between LEDBAT flows is instead depicted in Fig. 3, again for the high speed scenario $C = 10$ Mbps, showing that the dynamics depend on the precise values of the buffer size $B$ and of the sources start time gap $\Delta T$. Fig. 3 reports the sender window of the two LEDBAT flows. Let us start by considering the top plot, obtained for $(\Delta T, B) = (2, 40)$: in this case, the second flow starts before the first has started to create queue in the buffer. Then when the second flow starts, the queue rapidly builds up as well as the queue-delay, and the target is met soon. Yet the two flows are contributing differently to the delay: in fact the first, having started before, is able to achieve a larger congestion window and actually owns the biggest share of the queue.

Whenever the second flows starts after a $\Delta T$ large enough to allow the first one to create some queueing delay, a different dynamic is triggered. This is highlighted in the middle plot of Fig. 3, obtained for $(\Delta T, B) = (10, 40)$. In this case, the second flow senses a base delay which exceeds the true base delay, correctly measured by the first flow $t_B(t_1 = 0)$, by an amount of queuing delay $t_Q(t_2) = \tau_T$ due to the fact that the first flow has achieved its delay target. Therefore, the second flow sets a delay target corresponding to a queuing delay equal to $t_B(t_2) + \tau_T = t_B(t_1) + 2\tau_T$, and starts increasing its window immediately after $t = \Delta T$. At the same time, the first flow senses an increasing queuing delay and slows down its sending rate: the decrease continues until the first flow has reached the minimum rate, which happen slightly after $t = 20$ s in the case of figure.

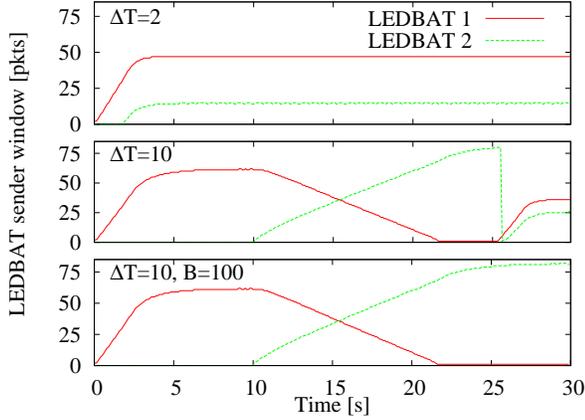

Fig. 3. LEDBAT vs LEDBAT: Time evolution of congestion window for different initial condition and late-comer advantage phenomenon

Then, dynamics depend on the specific buffer settings: when the buffer is not large enough to accommodate the queuing delay target of the second flow (i.e., B=40 < $2\tau_P$=41.6), the second flow rate grows so much to induce a loss on the bottleneck link, as can be seen around $t = 25$ s in the middle plot of Fig. 3. Though the loss implies a drop of the sending rate, it has the beneficial side effect of letting both sources to correctly measure the base delay, as the queue empties after the loss event. In a sense, the loss event resynchronizes the start of the flows, which then share more fairly the bottleneck bandwidth (as it can be seen for $t > 25$ s onward).

Conversely, whenever the buffer is large enough (e.g., B=100) to absorb the excess queuing delay introduced by the second flow, another phenomenon happens, as reported in the bottom plot of Fig. 3. Since no loss occurs, the second flow does not reduce its sending rate and is able to reach its queuing target. Once the target is reached, the second flow settles, leaving the first flow in starvation. This extremely unfair state may persist for possibly long time[1], raising the need to cope with this potentially serious unfairness problem.

### C. Side Effects of Slow-Start

We have seen that in the LEDBAT-LEDBAT interaction, the linear controller *alone* may get stuck in an unfair state during a relatively long time. Yet, comparing the middle and bottom plots of Fig. 3, we gather a very important observation: whenever a loss event happens, the competing flows may be able to re-establish fairness (at least to a certain degree).

In other words, a loss event resynchronizes the start of the flows, possibly draining the queue empty and thus allowing each flow to gather correct measures of the base delay. Extending this observation, it seems as though it is *necessary* for each LEDBAT flow to force a loss event at startup, so to gather a correct measure of the base delay: a simple, though intrusive, way to achieve this is to enable *slow-start*. As [7] lacks a precise description of the LEDBAT slow-start (which is only briefly mentioned as an optional feature for conservative LEDBAT implementations), we resort to standard TCP slow-start mechanism. In TCP, slow start is performed by initially setting ssthresh to $\infty$, performing an exponential window increase and then, in case of loss, setting ssthresh = cwnd/2 and cwnd=0: this process iterates until the window exceed ssthresh, in which case the slow-start phase ends.

We gauge the impact of slow-start on the network and user performance in terms of efficiency $\eta$, fairness $F$ and loss rate $L$. Notice that, although the precise evaluation of the impact of LEDBAT slow-start on VoIP/Gaming flows is outside the scope of this work, nevertheless we may gather an indirect observation of its impact by measuring the loss probability $L$.

As before, only two flows share the bottleneck and we consider i) the ideal case where neither TCP nor LEDBAT implement slow-start, ii) a more realistic case where both TCP and LEDBAT implement the same slow-start behavior. To examine late-comer situation, we neglect the case $\Delta T = 0$, since no fairness issues were observed in this case, and instead consider the start time of the second flow to be uniformly distributed in $\Delta T = U(0, 10)$ s, reporting the average of 100 simulation runs. For reference, we also consider the two values of $\Delta T \in \{2, 10\}$ s reported early in Fig. 3, and perform 10 simulation runs per each value of $\Delta T$ (jittering the start time of the second flow by a time lapse uniformly distributed in $[0, 0.1]$ s at each run). We now consider both the low-capacity $C_{ADSL} = 2$ and high-speed $C_{HS} = 10$ cases, and set the buffer size $B$ to values slightly above the bandwidth delay product and able to accommodate about twice as much as the delay target of LEDBAT flows. Simulation lasts for 300 seconds, and results refer to the time interval $[\Delta T, 300]$ s where both flows are active at the same time.

Results are reported in Tab. I. Top part of the table reports the TCP vs LEDBAT case, while LEDBAT vs LEDBAT is reported at the bottom. Left portion of the table refers to the case when no slow-start is used, while results obtained when slow-start is activated are reported on the right portion.

It can be gathered that simulation results confirm our intuition: the slow-start phase allows LEDBAT flow to reintroduce fairness on the LEDBAT vs LEDBAT case, while leaving the TCP vs LEDBAT case almost unchanged. For instance, notice that in the worst-case for the fairness metric (represented by $(C, B, \Delta T)=(10, 50, 10)$ where the behavior is similar to the one early reported in the middle plot of Fig. 3), the use of slow-start raises the LEDBAT vs LEDBAT fairness from $F = 0.53$ to $F = 0.99$. Even in the extreme case (not shown in the table) of a capacity $C = 2$ Mbps and a buffer $B = 100$ packets, i.e., and ADSL link with a very large buffer (about 500 ms), the fairness between two ledbat flows increases from $F = 0.57$ to $F = 0.77$ when slow-start is used (with a limited loss rate $L = 4 \cdot 10^{-3}$).

---

[1] Due to route change consideration, [7] requires the computation of a new minimum (which will break the persistence in the state and trigger further changes in the window dynamics) after about BASE_HISTO∈ [2, 10] minutes.

TABLE I
LINK UTILIZATION $\eta\%$, MEAN $\mu$ AND STANDARD DEVIATION $\sigma$ FAIRNESS $F$ AND LOSS RATE $L$. TCP VERSUS LEDBAT AND LEDBAT VERSUS
LEDBAT SCENARIOS, WITH/WITHOUT SLOW-START, FOR DIFFERENT CAPACITIES $C$, BUFFER SIZES $B$ AND TIME GAP $\Delta T$.

| Scenario | $C$ Mbps | $B$ Pkts | $\Delta T$ sec | Without Slow-Start | | | | | With Slow-Start | | | | |
|---|---|---|---|---|---|---|---|---|---|---|---|---|---|
| | | | | $\eta$ [%] | $F$ $\mu$ | $\sigma$ | $L$ $\mu$ | $\sigma$ | $\eta$ [%] | $F$ $\mu$ | $\sigma$ | $L$ $\mu$ | $\sigma$ |
| TCP LEDBAT | 2 | 10 | 2 | 99 | 0.60 | $6.5\cdot10^{-4}$ | $6.2\cdot10^{-3}$ | $9.4\cdot10^{-6}$ | 99 | 0.58 | $1.0\cdot10^{-3}$ | $1.5\cdot10^{-2}$ | $1.5\cdot10^{-3}$ |
| | | | 10 | 97 | 0.60 | $4.2\cdot10^{-3}$ | $6.2\cdot10^{-3}$ | $2.1\cdot10^{-5}$ | 94 | 0.58 | $2.6\cdot10^{-3}$ | $1.3\cdot10^{-2}$ | $9.7\cdot10^{-4}$ |
| | | | U(0,10) | 98 | 0.61 | $6.8\cdot10^{-2}$ | $6.2\cdot10^{-3}$ | $4.5\cdot10^{-4}$ | 98 | 0.60 | $4.5\cdot10^{-3}$ | $6.6\cdot10^{-3}$ | $4.1\cdot10^{-5}$ |
| | 10 | 50 | 2 | 99 | 0.53 | $1.1\cdot10^{-3}$ | $3.0\cdot10^{-4}$ | $1.3\cdot10^{-6}$ | 99 | 0.57 | $6.4\cdot10^{-3}$ | $1.2\cdot10^{-3}$ | $1.1\cdot10^{-4}$ |
| | | | 10 | 97 | 0.55 | $8.0\cdot10^{-4}$ | $3.1\cdot10^{-4}$ | $1.0\cdot10^{-8}$ | 97 | 0.58 | $6.8\cdot10^{-3}$ | $1.3\cdot10^{-3}$ | $1.1\cdot10^{-4}$ |
| | | | U(0,10) | 98 | 0.54 | $4.6\cdot10^{-3}$ | $3.0\cdot10^{-4}$ | $2.4\cdot10^{-6}$ | 98 | 0.55 | $1.8\cdot10^{-3}$ | $6.8\cdot10^{-4}$ | $3.8\cdot10^{-6}$ |
| LEDBAT LEDBAT | 2 | 10 | 2 | 99 | 0.70 | $1.2\cdot10^{-1}$ | $5.8\cdot10^{-5}$ | $3.8\cdot10^{-5}$ | 99 | 0.85 | $6.5\cdot10^{-2}$ | $7.1\cdot10^{-4}$ | $8.2\cdot10^{-6}$ |
| | | | 10 | 96 | 0.80 | $1.8\cdot10^{-1}$ | $4.8\cdot10^{-5}$ | $4.2\cdot10^{-5}$ | 96 | 0.83 | $5.8\cdot10^{-2}$ | $6.4\cdot10^{-4}$ | $5.7\cdot10^{-5}$ |
| | | | U(0,10) | 98 | 0.83 | $1.8\cdot10^{-1}$ | $3.8\cdot10^{-5}$ | $3.7\cdot10^{-5}$ | 98 | 0.83 | $1.0\cdot10^{-1}$ | $1.1\cdot10^{-3}$ | $2.3\cdot10^{-3}$ |
| | 10 | 50 | 2 | 99 | 0.73 | $4.4\cdot10^{-2}$ | - | - | 99 | 0.93 | $9.6\cdot10^{-2}$ | $4.3\cdot10^{-4}$ | $1.3\cdot10^{-8}$ |
| | | | 10 | 97 | 0.53 | $4.7\cdot10^{-4}$ | - | - | 96 | 0.99 | $2.6\cdot10^{-3}$ | $4.1\cdot10^{-4}$ | $2.0\cdot10^{-6}$ |
| | | | U(0,10) | 98 | 0.64 | $1.8\cdot10^{-1}$ | - | - | 98 | 0.96 | $8.3\cdot10^{-2}$ | $4.4\cdot10^{-4}$ | $5.9\cdot10^{-5}$ |

Concerning the loss rate, we expect slow-start to generate loss events only at the start of each connection: therefore, we expect the loss rate $L$ to be limited. From the table, we gather indeed that, despite the loss rate grows by about one order of magnitude when slow-start is enabled, nevertheless the absolute amount of losses is always very limited. In case only LEDBAT flows, with slow-start enabled, share the bottleneck, loss rate tops to about $L = 1 \cdot 10^{-3}$ in the worst case. Thus, it seems as though the impact on CBR VoIP/Gaming flows is likely to be negligible, although a more realistic evaluation is definitively needed (e.g., taking into account the VoIP codec and framing, the loss pattern, an higher number of LEDBAT flows with different arrivals, etc.).

## V. RELATED WORK

Two bodies of work are related to this study. On the one hand, there is a large literature on Internet congestion control algorithms, carried on with diverse tools such as simulation and modeling [13]–[19], or on fields measurement [21]–[23]. On the other hand, there are studies that focus on other important aspects of BitTorrent, that again exploits either theoretical analysis [24], simulation [25], [26] or measurements [27].

Motivated by the so called congestion collapse that hit a young Internet, distributed algorithms for the allocation of resources were invented and analyzed, starting from the seminal work on TCP [13]. A huge literature exists on the topic and, as a result, TCP comes in different flavors. However, most of this work focuses on ameliorating the performance of best-effort traffic, while the aim of LEDBAT is to achieve "lower" than best-effort performance. Under this light, closer to our work are [15]–[18] (although a linear decrease of the congestion window was already introduced in [19]). Lower priority is obtained by either adapting the *sender* window on the basis of loss rate [17] or delay measurements [15], [16], or by tuning the *receiver* window at the application layer [18]. For further details, we refer the reader to the respective publication or to [20] for their overview. Finally, it is worth mentioning that, related to this work are also studies that adopts a complementary approach, based on black-box experimental measurements, to unveil closed and proprietary congestion control algorithms of novel P2P systems (such as Skype [21], [22] or P2P-TV applications [23]).

Due to its recent evolution, previous work on BitTorrent [24]–[27] focused on complementary aspects to those analyzed in this work. In [24] a fluid model is used to determine the average download time of a single file. Simulation has instead been used to analyze and improve BitTorrent performance, as for instance in [25] and [26] where mechanism to prevent free-riding beyond tit-for-tat and a locality-aware peer selection mechanisms are proposed respectively. Finally, BitTorrent has been analyzed also through measurement studies such as in [27], where authors analyze the log of a BitTorrent tracker, examining flash-crowd effect popularity and download speed of a single file. However, due to BitTorrent very recent evolution, to the best of our knowledge no work focusing on the new congestion control protocol used for data dissemination has appeared yet.

## VI. DISCUSSION AND CONCLUSIONS

In this paper, we report on the evaluation of LEDBAT, a novel congestion control protocol for low-priority data transport, which aims at being friendly and non-intrusive toward other protocols such as TCP, VoIP and gaming, while at the same time being effective in exploiting the available resources.

By means of simulation, we illustrate interesting aspects of LEDBAT congestion window dynamics in simple scenarios. Concerning LEDBAT performance, our evaluation shows that:

- LEDBAT is able to achieve inter-protocol *friendliness* (i.e., yield to TCP) while being able at the same time to efficiently exploit the extra available resources.
- Inter-protocol *fairness* is maintained even in case of wrong parameter settings: indeed, when TARGET is too big with respect to the buffer size, LEDBAT degenerates into TCP, since it linearly increases the sender window to reach the target until drop happens.

- The linear controller *alone* is not sufficient to guarantee intra-protocol fairness: indeed, provided that buffer is large enough, a late-comer advantage may arise among LEDBAT flows.
- Intra-protocol fairness can be achieved provided that newcomer flows are given the chance to correctly measure the base delay, which can be accomplished in an uncoordinated and distributed fashion by simply using a *slow-start* phase.
- The latter observation also suggests that it may be necessary to make slow-start *mandatory* in the draft requirement; interestingly, slow-start happens to be necessary for its beneficial side effect on fairness, more than for efficiency reasons.

Though these preliminary results are interesting per se, nevertheless they only convey a limited view of the potential impact of a widespread adoption of LEDBAT in the Internet. First of all, simulation on a wider range of scenarios (e.g., heterogeneous RTT, multiple flows, impact on the QoE of VoIP traffic, comparison with other low-priority approaches, etc.) is needed in order to further refine the picture. Then, we believe that effort should be devoted also to modeling LEDBAT dynamics, in order to confirm simulation evidence with more theoretical findings. Finally, another interesting point concerns the empirical evaluation of the LEDBAT implementation in BitTorrent, which could be tackled by black-box measurement. Our future work intend to follow the above directions.

Besides, notice that the success of LEDBAT will be determined, first of all, by its user and their consensus: an important point in this regard concerns the degree of "low-priority" of LEDBAT, or the "right" fairness balance. Indeed, while it is very important that LEDBAT avoids harming interactive traffic (e.g., Web, Mail, VoIP, Gaming), it is less reasonable for LEDBAT to yield to non-interactive TCP traffic as well (e.g., long FTP transfer, TCP transfers of other P2P applications, etc.). The ability of LEDBAT to yield to interactive traffic is indeed a good incentive for user adoption, as it improves user experience concerning troubles induced by self-congestion at the access. Yet, while users will surely welcome this LEDBAT feature, they will be less inclined in tolerating the this very same friendliness toward the Internet traffic of other users – especially in case this could translate in poorer P2P performance for themselves. In the case of BitTorrent, an important question that remains open is, for instance, how much the donwload time will be degraded by the adoption of this new congestion control protocol.

Overall, our results confirm that LEDBAT has an undoubted and promising appeal to become a very useful Internet building block – and moreover its implementation in one of the most popular P2P application already constitutes a good starting point to achieve this goal.

ACKNOWLEDGEMENT

This work has been funded by the Celtic project TRANS.